\documentclass[preprint]{aastex}

\usepackage{graphicx}

\newcommand{\be}{\begin{equation}}
\newcommand{\ee}{\end{equation}}
\newcommand{\nn}{\mbox{} \nonumber \\ \mbox{} }
\newcommand{\ba}{\begin{eqnarray}}
\newcommand{\ea}{\end{eqnarray}}
\newcommand{\om}{\omega}
\newcommand{\Alfven}{Alfv\'{e}n }

\newcommand\etal{\textit{et al.\ }}
\newcommand\eg{\textit{e.g.\ }}

\newcommand\ie{\textit{i.e.\ }}

\newcommand{\Chandra}{{\it Chandra\, }}
\newcommand{\pd}[2]{{{\partial #1}\over {\partial #2}}}
\newcommand{\green}{\mathcal G}

\begin{document}

\date{}  
\title{The origin of non-thermal X-ray filaments and TeV emission in young SNRs}
\author{M. Lyutikov\altaffilmark{1}}
\affil{Physics Department, McGill University, 3600 rue University,
Montreal, QC,\\Canada H3A 2T8}
\altaffiltext{1}{Canadian Institute for Theoretical Astrophysics,\\ 60 St. George, Toronto, Ont,  
M5S 3H8, Canada}
\author{M. Pohl}
\affil{Department of Physics and Astronomy,
Iowa State University\\
Ames, Iowa 50011-3160, USA}
\email{lyutikov@physics.mcgill.ca; mkp@iastate.edu}

\begin{abstract}
At the early, ejecta dominated, stage of supernova  remnant (SNR) expansion
a fraction of the swept-up circumstellar 
magnetic field is dynamically compressed to approximate equipartition
at the contact discontinuity separating the SN progenitor's wind
(or the ISM) and the ejecta. We propose that the thin non-thermal
X-ray filaments observed by the Chandra satellite in several
young SNRs are associated with such ``pile-up'' of the magnetic field. 
{We use a one-dimensional diffusion-convection transport equation 
to describe the propagation of non-thermal electrons near the contact discontinuity of a young
SNR and to calculate spatially resolved emission spectra in the X-ray and TeV bands. 
The results suggest that} the
high-energy electrons {are possibly accelerated  at the forward shock, and
emitting efficiently only when they diffuse into regions of high magnetic field
near the contact discontinuity.
Much more likely, though, is that they are locally accelerated at the contact 
discontinuity,
in which case the acceleration cannot be related to Fermi-type processes and should
occur due to other plasma mechanisms.}
As a consequence,   the forward shock
 in young SNRs is inconspicuous
 and 
often
unobservable, similar to that in the Crab nebular.
\end{abstract}

\section{Introduction}

A  number of young supernova remnants (SNRs)  produce
non-thermal  X-ray emission (SN 1006, Koyama \etal 1995;
Cas A, Allen \etal 1997; RX J1713.7-3946, Koyama \etal 1997;
RCW 86, Borkowski \etal 2001, Bamba  \etal 2000, Rho \etal 2002;
G266.2-1.2, Slane \etal 2001).
Recent high resolution Chandra observations of these SNRs
indicate that a large fraction of the non-thermal X-ray emission is produced in very thin
structures in the outer regions of the remnants
(Cas A, Gotthelf \etal 2001; SN 1006,  Bamba \etal 2001;  RX J1713.7-3946,
Uchiyama \etal 2003). The filament thickness is smaller than one tenth of a parsec, 
and possibly as small as one hundredth of a parsec. 

In addition to X-rays, most prominent non-thermal SNRs show TeV-scale
$\gamma$-ray  emission (SN 1006, Tanimori \etal 1998;  RX J1713.7-3946, 
Muraishi \etal 2000; Cas A, Aharonian \etal 2001a;
see also the review by  Vink 2004). Though the CANGAROO  TeV $\gamma$-ary  detections
of SN~1006 and  RX~J1713.7-3946 are not confirmed to date, 
the overlap between the two populations is
notorious: all TeV SNRs show non-thermal  X-ray emission.
The origin of the
TeV emission is been currently debated. It has been attributed either to
IC  scattering of CMB photons by high energy electrons
 or to the decay of  $\pi^0$
during interaction of the cosmic rays accelerated at the forward shock
with ambient material (Berezhko, Ksenofontov, \& V\"olk 2002, 
Berezhko, P\"uhlhofer, \& V\"olk 2003).

The overall properties of the non-thermally emitting  X-ray SNRs are quite different
(Pannuti \etal 2003).  SN 1006 was produced by a Type Ia SN, while
RX J1713.7-3946 and Cas A are the result of core-collapse Type II SNe. 
SN 1006 is  expanding into a low density environment, while
Cas A and RX J1713.7-3946  are expanding into a wind blown cavity
(Chevalier \& Oishi 2003; Slane \etal 2001). We suggest that the uniting
property  is that these are young {\it pre-Sedov} SNRs (also non-plerionic). 
The  reason why this may influence the non-thermal X-ray emission is that
in the  {\it pre-Sedov} phase the
 magnetic field is  compressed  to near equipartition values 
 at the contact discontinuity separating the ejecta
and the circumstellar medium.
This may create suitable conditions for an efficient electron acceleration
{\it  at the  contact discontinuity}.

The extreme narrowness of the filaments poses several problems for their interpretation.
Conventionally, electrons are assumed to be accelerated at the forward shock by Fermi
acceleration, in a way similar to cosmic rays. The non-thermal  X-ray emission
in the SNRs is generally thought to be synchrotron emission from
electrons that have been  accelerated to hundreds of TeV at the forward shock
(see, though, Vink and Laming (2003)  for a nonthermal bremsstrahlung interpretation).
 However,
there are a number of problems that one encounters when trying
to associate the narrow non-thermal filaments with acceleration at the forward shock.
\begin{itemize}
\item
If the magnetic field is just the compressed ISM field, then the
synchrotron decay times are {usually too long to account for the narrow filaments}.
To reproduce thin filaments, the 
required  magnetic field should be very high, of the order of equipartition with
$B\sim 0.1-1$~mG,
so that X-ray emitting electrons  would lose most of their energy 
before they are advected or diffuse away  from the filaments.
\item The acceleration of electrons at the forward shock should produce
a precursor {in the upstream region} of the shock. In the cleanest case of 
SN 1006 this is not seen
with a very low upper limit (Long \etal 2003).
\item  
In RCW 86 the detection of a strong Fe $K_\alpha$ line by 
Rho \etal (2002) showed that it cannot be produced by thermal emission
from a cosmic-abundance plasma, but it can be produced
by Fe-rich  ejecta.
Similar conclusions has been reached by Bamba  \etal (2003) for the case of SN 1006.
(Please note again that SN 1006 was caused by a Type Ia SN.)
These observations seem to exclude acceleration at the forward shock, for there
is no heavy element enhancement around the forward shock.
\item The acceleration efficiency required to explain the very high energy synchrotron
cut-offs (above 10 keV) is too high for shock drift acceleration 
(\eg Uchiyama \etal 2003 in case
of SNR RX J1713.7.3946).
\footnote{
In the framework of the diffusive shock acceleration model, the synchrotron cutoff energy
is set by the condition acceleration rate = synchrotron loss rate.
The diffusive shock acceleration time 
 is
$t_{acc} \sim 10 \kappa/V^2$, where $\kappa= \eta r_L c/3$ is the diffusion coefficient
in the upstream region, and V is the upstream velocity of the shock,
$\eta \geq 1$ is the gyrofactor. The maximum energy
of synchrotron emission is then
$ \epsilon_X = 2 (V/2000 km s^{-1}) ^{2} /\eta $ keV.
Thus, it is very hard, if not impossible, to reproduce observed non-thermal
emission reaching  in some cases 10 keV (Uchiyama \etal 2003).}
\item 
Attempts at the broadband (X-ray through TeV $\gamma$-ray)
 modeling of SNR emission has encountered
difficulties. This is exemplified by the best studied case of RX J1713.7-3946. 
Initially, Enomoto \etal (2002) concluded that the TeV spectrum is 
a good match to that predicted by pion decay, and cannot be explained by other mechanisms.
This was questioned by Reimer \& Pohl (2002) and Butt \etal (2002),
who concluded that the pion-decay model adopted by Enomoto
\etal (2002) is in conflict with the existing data.
They argued that, though  RX J1713.7-3946 is accelerating electrons to super TeV
energies, no convincing evidence for hadron acceleration exists to date.
Finally, Pannuti \etal (2003) found that neither non-thermal bremsstrahlung
nor neutral pion decay can adequately describe the  TeV emission. 
On the other hand, the  synchrotron-inverse Compton model can
reproduce the data with magnetic field 150  $\mu$G, but the
 derived ratio of volumes for TeV emission and X-ray emission (approximately 1000)
was deemed too large (see also Lazendic \etal 2004).
\end{itemize}

In attempt to resolve these contradictions, Berezhko, Ksenofontov, and V\"olk (2002; 2003) 
suggested that very efficient acceleration of cosmic-ray
nucleons may indeed cause near-equipartition magnetic
fields on account of streaming instabilities near the forward shock
(McKenzie \& V\"olk 1982; Lucek \& Bell 2000; Bell \& Lucek 2001).
In  this case,  the relative fluctuation amplitudes of the magnetic field
have to be  much larger
than unity, $\delta B /B \gg 1$, {the cosmic-ray diffusion should be in the Bohm limit,}
and the total energy density in
cosmic rays and in the magnetic field are of the order of equipartition.
If the magnetic field is that high and is homogeneous over the remnant, the
total magnetic energy would be similar to the expansion kinetic energy of the remnant. 
These are seriously constraining conditions. {
Also, the non-detection of TeV-scale gamma-ray emission
from shell-type SNRs in a high-density environment does not support the notion of 
cosmic-ray nucleons acceleration to energy densities at the equipartition level 
(Buckley \etal 1998; Aharonian \etal 2001b, 2002).}

In this paper we investigate an alternative possibility that 
non-thermal X-ray filaments are associated with the contact discontinuity separating 
ejecta and circumstellar material.
We point out that there is a natural way to produce small scale
near-equipartition magnetic fields in young (pre-Sedov) SNRs due to the
{\it dynamical compression of magnetic fields at the contact discontinuity},
see Section \ref{Bf}. 
 This possibility has already been noted by
Kulsrud \etal (1965), who considered the expansion of the ejecta
in a medium of constant density and constant magnetic field, and  by
Rosenau \etal (1976) (see also Lee and Chen 1968), who considered {the propagation of
magnetohydrodynamic shocks into an ideal gas permeated by a current-free}
toroidal magnetic field (for a relativistic wind
this has been done by Lyutikov 2002). 

While the progenitors of core-collapse supernovae blow bubbles into the ambient medium,
their wind carries magnetic flux with it.  After the explosion,
the supernova ejecta propagates through the wind-blown bubble, sweeping magnetic flux
with it. A fraction of the swept-up magnetic flux piles up the contact discontinuity, where the
magnetic field reaches equipartition {with the thermal flow}.
Similarly,  {the reverse shock sweeps up}
magnetic flux of the  ejecta, part of which piles up on the contact discontinuity. 

{The distinct possibility that the thin X-ray} filaments are associated with the
contact discontinuity may resolve the inconsistencies of the forward shock
acceleration model listed above.
\begin{itemize}
\item Since there is no flow of material across
the contact discontinuity, there is no problem with the advection of accelerated electrons. 
Our estimates
of the diffusion rate show that the diffusion length in one
synchrotron decay time is short enough to be consistent with 
the thickness of filaments. 
\item High values of the magnetic field
decrease electron diffusion away from the sheath. In addition,
 since the region of high magnetic
field is small, the synchrotron emissivities are small outside of the sheath.
\item Near the contact discontinuity the background plasma consists of both circumstellar medium
and ejecta, which is enriched with heavy elements. 
Acceleration and the associated thermal heating
may occur on both sides of the contact discontinuity (in the ejecta and in the
circumstellar material). 
\item It is feasible, though we do not demonstrate it, that plasma acceleration 
mechanisms are more efficient than standard shock acceleration.
\item A large
ratio of TeV to synchrotron emitting volumes is a natural consequence of 
the magnetic field compression.
\end{itemize}

\section{Magnetic fields  of young SNRs }
\label{Bf}

The expansion of SN ejecta into the circumstellar medium has been the subject
of numerous studies (see, \eg, Truelove \& McKee 1999 for the latest review),
 so that the hydrodynamic behavior is well understood.
The dynamic effects of the magnetic field have been largely ignored, though. 
For the typical magnetic field this is a justified  assumption
{\it except}  near the contact discontinuity separating the  shocked ejecta 
and the circumstellar medium.

The dynamics of the shocked wind strongly depends on whether the explosion is
in the Sedov phase or in the ejecta-dominated phase,
when the piled-up mass is still smaller than the ejecta mass, \ie at times $
t \leq E^{-1/2} M_{ej}^{5/6} \rho_0^{-1/3} $
(\eg Truelove \& McKee 1999), where  $E$ is the energy of the explosion, $M_{ej} $  is
the ejecta mass, and $ \rho_0$ is the external density. 
In the Sedov phase, most of the shocked circumstellar material
and, due to the "frozen-in" condition, the magnetic field
is concentrated in  a thin layer near the shock surface.
On the other hand, in the  ejecta-dominated phase the presence of the
contact discontinuity between the  circumstellar material and ejecta
changes the dynamics considerably. In particular, for a wide variety
of ejecta and wind profiles the plasma density formally diverges
on the contact discontinuity. On account of the frozen-in condition, this means that the magnetic field
diverges as well (Kulsrud \etal 1965; Rosenau 1975; Rosenau \& Frankenthal 1976).
The reverse shock propagating in the ejecta leads to a similar, perhaps even more 
prominent, effect (Hamilton \& Sarazin 1984).  
{\it Thus, no matter how weak the preshock magnetic field
is, there is always a thin boundary layer near the contact discontinuity
where the magnetic pressure is comparable to or even dominates over the
kinetic pressure.} The thickness of the layer depends on the wind and the ejecta
magnetic field.  Exact solutions
(\eg Rosenau 1975) indeed show that on the contact discontinuity the
pressure is provided solely by the magnetic field, and the kinetic pressure vanishes.
\footnote{A similar effect occurs  for the
interacting winds case (Emmering \& Chevalier 1987)}

Next we outline the dynamics near the contact discontinuity (for details see Rosenau \& Frankenthal 1976).
Assume that the ejecta is propagating through a cold  supersonic magnetized
wind with a constant velocity $v_0 $
and a decreasing mass density $\rho \propto r^{-2}$ (this is expected for 
a supersonically moving pre-supernova wind). Let  
the speed of the forward shock be $v_s$.
The assumption of a cold plasma implies that the sonic
Mach number is infinity.
In the wind the magnetic field is expected to be dominated
by the toroidal component $B_\phi \propto 1/r$, so that the \Alfven
velocity, $v_A = B/ \sqrt{ 4 \pi \rho}$, remains constant.
We will parameterize the strength of the magnetic field
by the \Alfven Mach number
$M_A= (v_s - v_0)/v_A \sim  v_s /v_A$, where we assume $v_s \gg v_0$.
Note that in  case of  the Solar wind the \Alfven velocity is somewhat smaller, but
comparable to the terminal velocity $v_A \leq v_0$.
After the passing of the forward shock the
magnetic field and the particle density are compressed by the factor
$(\gamma+1)/(\gamma-1) = 4$. At the same time the post-shock pressure
is  $p \sim \rho v_s^2 \gg B^2/4 \pi $, so that the post-shock plasma is strongly
dominated by the kinetic pressure.
Thus, {\it in the bulk} of the shocked flow dynamic effects of magnetic field
are mostly negligible (see, though, Chevalier \& Luo 1994), but it cannot be neglected
in a narrow sheath near the contact discontinuity.

To estimate the width of the magnetized sheath in the swept-up
 circumstellar  medium we need to specify
the motion of the shock and can then use the solutions
of Rosenau \& Frankenthal (1976).
For simplicity, we assume here that the SNR is well in the
ejecta-dominated regime. In this case, the velocity of the shock
is constant $v_s = {\rm  const}$, while the structure of the shocked wind
is self-similar, depending only on  $ \xi  = r/R_S$, where $R_S$ is the shock position
at a given time. The contact discontinuity is located at $\xi_{CD} \simeq 0.78$.
For large \Alfven Mach numbers $M_A \gg 1$ the flow dynamics far from the
contact discontinuity follows the hydrodynamical case. In the vicinity of the contact discontinuity the 
{dynamical behavior} is complicated, though.
In {the absence of a} magnetic field the kinetic  pressure remains
constant, while the density tends to infinity and the temperature to zero. {
For an adiabatic index $\gamma =5/3$ the three quantities obey the following scaling relations
in $\hat{\xi} = \xi - \xi_{CD} \ll 1$:}
\be
p=  {\rm const.}, \hskip .3 truein 
\rho \sim \hat{\xi}^{-4/9} \rightarrow \infty
, \hskip .3 truein
T \sim \hat{\xi}^{4/9}
\ee
If the flow carries a small magnetic field,
then the frozen-in condition requires
\be
B  \sim \rho \sim \hat{\xi}^{-4/9} \rightarrow \infty
\ee
Therefore, however small the magnetic field is, it
 becomes dynamically important near the contact discontinuity.
Thus the flow may be separated into a bulk flow, where the magnetic field is 
dynamically not important,
and a magnetized sheath where its pressure is comparable to the kinetic pressure.
In the magnetized sheath
\be
p \sim \hat{\xi}^{4/3} , \hskip .3 truein
\rho \sim {\rm const.}
, \hskip .3 truein
B=  {B_0 } = {\rm const.}
, \hskip .3 truein
T \sim  \hat{\xi}^{4/3}
\ee
Since the magnetic field on the CD should transport all the momentum from the ejecta,
$B_0$ will be of the order of the equipartition pressure.

The width of the magnetized layer, $\Delta R_{\rm sheath} $, is
\be
\Delta \zeta \simeq {\Delta R_{\rm sheath} \over R} \simeq  M_A^{-9/4}
\ee
It  may be shown that only a small fraction $\sim M_A^{-5/4}$
 of the total swept-up flux  is concentrated near the
contact discontinuity. Most of the swept-up flux is still near the shock surface, {for
in the ejecta-dominated phase the self-similar solutions for the forward shock
 have similar properties as in the Sedov phase.
Though most of the magnetic flux is distributed near the shock surface,
the magnetic field near the contact discontinuity is $\sim M_A$ times higher, so that
for a homogeneous electron distribution
the synchrotron emissivity is  $M_A^2$ times higher near the contact discontinuity. 
The synchrotron emission produced near the contact discontinuity can thus dominate the total radiation yield,
in particular if one observes the sheath edge-on.}

To estimate the physical size of the magnetized sheath one must know the
parameters of the pre-supernova wind, a highly uncertain and varying quantity.
For a Type II SNR remnant propagating into a wind blown cavity
it will depend on whether the forward shock is 
propagating in the fast, $\sim 1000$ km /sec WR wind, in the preceding slow wind of the LBV, 
or in the progenitor stage wind, usually that of an
O-star. In all cases the magnetic fields in the wind are virtually unknown
(\eg Cassinelli 2001).  For a qualitative estimate we assume that the 
wind is propagating at a few hundred kilometers per second, and the
shock velocity of the supernova is several thousand  kilometers per second. 
To estimate the magnetic field we assume that the 
\Alfven velocity at the sonic point of the wind is of the order of the sound speed
and that the terminal velocity of the wind is typically
only a few times larger than the velocity at the  sonic
point. Then the \Alfven Mach number is 
 $M_A \sim 10-30$ and
$\Delta \zeta \simeq 10^{-2} - 10^{-3}$. 

Similarly, it is possible to  estimate the thickness of the 
magnetized sheath in the ejecta using the self-similar solutions
of Hamilton \& Sarazin (1984). Unfortunately, it crucially depends on the 
density profile of the ejecta (see, \eg, Truelove \& McKee 1999) and also on the even 
more uncertain value of the magnetic field in the ejecta. Nevertheless, the pile-up of the 
ejecta magnetic field may be more important than that of the circumstellar medium,
for downstream of the reverse shock a larger fraction 
of the shocked material may pile up on the contact discontinuity, whereas downstream of the forward 
shock most of the material is still near the shock.  
For example, in  the self-similar solutions
of Hamilton \& Sarazin (1984), $\rho \propto \hat{\zeta}^{-6/5}$ and
the thickness of the magnetized sheath in the ejecta is
\be
\Delta \zeta_{ej} \sim M_{A, ej}^{-6/5},
\ee
where $ M_{A, ej}$ is the \Alfven Mach number of the reverse shock, so that
almost all of the magnetic flux, that passed through the reverse shock,
piles up on the contact discontinuity.

\section{Modeling non-thermal  X-ray emission}

\subsection{Where are the high-energy particles accelerated?}

{Having established that a thin sheath of high magnetic field strength should exist at the
contact discontinuity, we will now discuss possible processes of particle acceleration and the locations thereof,
for which two possibilities seem to exist.} First, relativistic electrons can
be accelerated at the forward shock, but they would not produce high-energy
emission efficiently, for the magnetic field is fairly low. 
As the accelerated electrons are advected downstream and diffuse through the shocked plasma,
they will encounter regions of high magnetic field at the contact discontinuity,
where they may produce non-thermal X-rays
and loose most of there energy. {Scenarios invoking a non-linear magnetic field amplification 
through streaming instabilities of cosmic rays, that have been produced by diffusive shock
acceleration at the forward shock (Berezhko, Ksenofontov, \& V\"olk 2003), are
apparently disfavored on account
of the X-ray spectrum of the filaments (Aharonian \& Atoyan 1999, Uchiyama \etal 2003). }

Secondly, the electrons may be accelerated
right in the high magnetic field regions near the contact discontinuity. In this case,
the acceleration could not be due to shock acceleration,{ for there is no jump in parallel
velocity at the contact discontinuity.}
Alternatively, the particle acceleration may be due to plasma instabilities that
develop in a thin magnetized sheath. This, in principle,
can be much more efficient. 
{
In this section we consider these two  alternatives.  We construct a model of 
propagation and synchrotron emission of electrons in an inhomogeneous magnetic and velocity
fields that should be present near the contact discontinuity. 
We  find that if electrons are accelerated far from the contact discontinuity, typically too few electrons
would diffuse into the  high field region near the contact discontinuity to produce bright narrow features,
unless the diffusion coefficient is very high.

Before doing detailed calculations let us first make a simple estimate 
whether  the size of the thin filaments is consistent with
local acceleration. Let's assume that electrons diffuse out of
the high  magnetic field
regions due to Bohm-type diffusion perpendicular to the magnetic field.
Let the diffusion coefficient be parameterized as
\be
\kappa_\perp = \lambda r_L c
\ee
where $r_L = \gamma c/\om_B$  is the relativistic Larmor radius
and $ \lambda \leq 1$ (for  Bohm  diffusion $\lambda=1$).
Then the diffusion scale  in one  synchrotron decay time, $\tau_c$, is
\be
l_{\rm diff} \simeq  \sqrt{ \lambda r_L \tau_c c}
\ee
Using the observed size of the filaments we can make an estimate
of the magnetic field strength assuming that the filament thickness is due to diffusion
{and energy losses}.
\be
\om_B \simeq \frac{c}{l_{\rm diff}}\,\sqrt[3]{6\pi\,\lambda\,{{l_{\rm diff}\over {r_e}}}}
\ee
where $r_e$ denotes the classical electron radius.
Note, that this estimate is independent of particle energy, so that, if taken at face
value, this would  imply that all emission (from radio to TeV) should be confined
to the filament. 
For the magnetic field we find
\be
B \simeq (150\ {\rm \mu G})\ \lambda^{1/3} 
\,\left({{l_{\rm diff}}\over {0.1\ {\rm pc}}}\right)^{-2/3}
\ee
For a given parameter $\lambda \le 1$ this gives a lower limit to the magnetic field, for
the thickness of filaments can be  determined not only by diffusive escape,
as would have been the case if the
acceleration occurred in an infinitely thin region, but also by the physical size
of the acceleration region itself.}

\subsection{Propagation of high-energy electrons}

{The X-ray filaments observed with \Chandra satellite
 in SNRs are very thin compared 
with the radius of the
remnants, and therefore we may treat the propagation of electrons as a one-dimensional problem.
If the contact discontinuity propagates outward at a constant velocity, it is 
convenient to use a comoving spatial coordinate, $z$, such that $z=0$ marks
the position of the contact discontinuity at all times. We approximate the energy loss by adiabatic
expansion in the spherical outflow by a catastrophic loss term
\begin{displaymath}
{\rm div}\,V=
{1\over {r^2}}\,\pd{}{r}\left(r^2\,V\right)\bigg\vert_{r=R_{\rm CD}}\simeq
2\,\frac{V_{\rm cd}}{R_{\rm cd}}
\end{displaymath}
\begin{equation}
\Rightarrow \ -\pd{}{E} \left[\frac{1}{3}\,{\rm div}\,V\;E\,N\right]
\simeq -\pd{}{E} \left[\frac{2}{3}\,\frac{V_{\rm cd}}{R_{\rm cd}}\,E\,N\right]
\ \longrightarrow\ \frac{N}{\tau_{\rm ad}}=\frac{2}{3}\,\frac{V_{\rm cd}}{R_{\rm cd}}\,N
\label{lp-eq1}
\end{equation}
For young SNRs like SN~1006 the adiabatic loss time is $\tau_{\rm ad}\approx 10^{11}\ $sec, longer
than the age of the remnant. In the case of RX~J1713.7-3946 the distance may be either 
$d\simeq 6\,{\rm kpc}$ (Slane \etal 1999), implying $\tau_{\rm ad}\approx 10^{12}\ $sec and an age a
factor of three less, or $d\simeq 1\,{\rm kpc}$ (Koyama \etal 1997), in which case 
$\tau_{\rm ad}\approx 1.8\cdot 10^{11}\ $sec and the age is a
factor of six less than that. Thus the adiabatic cooling associated with the spherical
expansion of the remnant is probably unimportant in the present phases of expansion for the remnants.
On the other hand, the  synchrotron loss time scale for synchrotron emission at an X-ray energies is
\begin{equation}
\tau_{\rm syn} \simeq 
(2\cdot 10^{10}\ {\rm sec})\,\left({B\over {20\ {\rm \mu G}}}\right)^{-1.5}
\,\left({{E_X}\over {4\ {\rm keV}}}\right)^{-0.5}
\label{lp-eq1a}
\end{equation}
so that continuous energy losses by synchrotron emission will not only be important near the contact discontinuity,
where the field strength can be very high, but also in the regions toward the forward and reverse shocks.  

Under these conditions
the dynamics of very high-energy electrons near the rim of a supernova remnant can be reasonably
well described by a one-dimensional continuity equation for the isotropic differential number density
of relativistic electron, $N(E,z)$, which incorporates the effects of diffusion, convection, 
adiabatic deceleration, and energy loss terms (Owens \& Jokipii 1977)
\begin{equation}
-\pd{}{z}\left[D(E,z)\pd{N}{z} - V(z)N\right]
-\pd{}{E} \left[\left(\frac{1}{3}\,{{dV}\over {dz}}\,E -\beta (E,z)\right)N\right]=Q(E,z)
\label{lp-eq2}
\end{equation}
with appropriate boundary conditions for $N(E,z)$. In Eq.\ref{lp-eq2} $D(E,z)$ is the scalar 
diffusion coefficient, $V(z)$ is the convection velocity of cosmic-ray electrons relative 
to the contact discontinuity, $Q(E,z,t)$ represents the source distribution, 
and $\beta(E,z)=\dot E$ is the rate of continuous energy losses. 
We assume that synchrotron emission is the dominant energy loss process.
Then 
\begin{equation}
\beta (E,z) =-E^2\,b(z)
\label{lp-eq5}
\end{equation}
where the spatial function $b(z)$ is proportional to the square of the magnetic field strength,
$B^2$. The diffusion coefficient is likely near the Bohm limit, for the freshly accelerated,
streaming cosmic rays very efficiently excite Alfv\'en waves (Wentzel 1974;
Skilling 1975a; 1975b; 1975c).
The diffusion coefficient will therefore decrease as $B$ increases. It thus seems appropriate to 
set
\begin{equation}
D(E,z) = D(E)\,d(z) = D_0\,E^a\,d(z)\, , \qquad 0\le a <1
\label{lp-eq6}
\end{equation}
with the scaling relation
\begin{equation}
d(z)\,b(z)=\alpha={\rm const.}
\label{lp-eq7}
\end{equation}
For $a\rightarrow 1$ we can recover the Bohm limit for the diffusion coefficient. 
Eq.\ref{lp-eq2} can then be rewritten in terms of the new spatial coordinate
\begin{equation}
\mu =\int^z \frac{ds}{d(s)}
\label{lp-eq8}
\end{equation}
as
\begin{equation}
-\pd{}{\mu}\left[D(E)\,\pd{N}{\mu} - V(\mu)N\right]
-\pd{}{E} \left[\left(\frac{1}{3}\,{{dV}\over {d\mu}}\,E +\alpha\,E^2\right)\,N\right]=d(\mu)\,Q(E,z)=
Q(E,\mu)
\label{lp-eq9}
\end{equation}
where the revised source density $Q(E,\mu)$ is now differential in $\mu$ instead of $z$. $N(E,\mu)$,
however, is still differential in $z$. For the case that the convection velocity relative to the contact discontinuity
is linear in the new coordinate $\mu$,
\begin{equation}
V(\mu) =3\,V_1\,\mu
\label{lp-eq10}
\end{equation}
The Green's function, $\green$, which satisfies
\begin{equation}
-\pd{}{\mu}\left[D(E)\,\pd{\green}{\mu} - 3\,V_1\,\mu\,\green\right]
-\pd{}{E} \left[\left(V_1\,E +\alpha\,E^2\right)\,\green\right]=\delta\left(E-E^\prime\right)\,
\delta\left(\mu-\mu^\prime\right)
\label{lp-eq11}
\end{equation}
for homogeneous boundary conditions at infinity, 
can be found in the literature (Lerche \& Schlickeiser 1981, see their Eq. 72) as
\begin{eqnarray}
\green&=&{{\Theta \left(E^\prime-E\right)}\over {2\,\sqrt{\pi}\,\left(V_1\,E +\alpha\,E^2\right)}}\,
{{\exp\left(3\,V_1\,\tau(E)\right)}\over 
{\left[\int_{\tau(E)}^{\tau(E^\prime)} dx\ D(x)\,\exp(6\,V_1\,x)\right]^{0.5}}}\,\nonumber \\
& &\times\ \exp\left(-{{\left[\mu\, \exp\left(3\,V_1\,\tau(E)\right)-
\mu^\prime\, \exp\left(3\,V_1\,\tau(E^\prime)\right)\right]^2}
\over {4\, \left[\int_{\tau(E)}^{\tau(E^\prime)} dx\ D(x)\,\exp(6\,V_1\,x)\right]}}\right)
\label{lp-eq12}
\end{eqnarray}
where 
\begin{equation}
\tau(E)=\int^E {{du}\over {V_1\,u + \alpha\,u^2}}
\label{lp-eq13}
\end{equation}
\begin{displaymath}
D(\tau)=D\left[\tau(E)\right]=D_0\,E^a
\end{displaymath}
and $\Theta(x)$ is the Heaviside function.

The general properties of this solution have been outlined in Lerche \& Schlickeiser (1982)
and Pohl \& Schlickeiser (1990). Here we will use a specific spatial profile for the magnetic field
strength (and for the diffusion coefficient) to describe the propagation of high-energy electrons near
to and in the pile-up region around the contact discontinuity of a young SNR. Though the pile-up of the magnetic field 
appears to be a general property of the systems, the calculations of Rosenau \& 
Frankenthal (1976) suggest that the spatial profile of the magnetic field strength near the contact discontinuity 
depends somewhat on the Alfv\'enic Mach number of the outflow and on the density profile 
of the upstream medium. We may therefore assume a simple mathematical profile
\begin{equation}
d(z)= {{1+\left({z\over{z_1}}\right)^2}\over {1+\left({z\over{z_0}}\right)^2}}\ ,\quad z_1\ll z_0
\label{lp-eq14}
\end{equation} 
corresponding to a variation of the magnetic field strength
\begin{equation}
B(z)\propto \sqrt{b(z)}= \sqrt{\alpha\,
{{1+\left({z\over{z_0}}\right)^2}\over {1+\left({z\over{z_1}}\right)^2}}}
\label{lp-eq15}
\end{equation} 
The magnetic field is thus assumed to be constant at large distances from the contact discontinuity, $z \gg z_0$.
It would increase approximately $\propto z^{-1}$ for $z_1\ll z\ll z_0$, and is constant again 
close to the contact discontinuity at $z \ll z_1$ with a value a factor of $z_0/z_1$ higher than far away from the contact discontinuity.

Then the new spatial coordinate, $\mu$, scales as
\begin{equation}
\mu=\left({{z_1}\over {z_0}}\right)^2\,z +\left[1-\left({{z_1}\over {z_0}}\right)^2\right]
\,z_1\,\arctan\left({{z}\over {z_1}}\right)
\label{lp-eq16}
\end{equation}
The convection velocity, $V$, is then linearly increasing as $V(z)=3\,V_1\,z$ for $z\ll z_1$. 
For $z\gg z_1$ the velocity gradient continuously decreases until it turns constant again for
$z\gg z_0$, where $dV/dz =3\,V_1\,z_1^2/z_0^2$. This velocity profile is in accord with the
findings of Rosenau \& Frankenthal (1976).

For ease of exposition we will assume the sources of electrons to be located either
at the contact discontinuity (case A) or far away from it (case B). The source terms for these two situations are
\begin{equation}
Q_A =q_0\,\delta(\mu)\, E^{-p}\,\Theta (E_{\rm max} -E) 
\label{lp-eq17}
\end{equation}
and
\begin{equation}
Q_B =q_0\,\delta(\mu-\mu_s)\, E^{-p}\,\Theta (E_{\rm max} -E) 
\label{lp-eq18}
\end{equation}
In the high-energy limit $\alpha E\gg V_1$, \ie synchrotron energy losses dominate, 
the differential number density of electrons then is 
\begin{equation}
N ={{q_0\,\sqrt{1-a}\,E^{-p-{{1+a}\over 2}}}\over {2\,\sqrt{\pi\,D_0\,\alpha}}}\,
\int_1^{{E_{\rm max}}\over E} dx\ {{x^{-p}}\over \sqrt{1-x^{a-1}}}\,
\exp\left[-\,{{(1-a)\,\alpha}\over {4\,D_0\,E^{a-1}}}\,{{(\mu-\mu_s)^2}\over {1-x^{a-1}}}\right]
\label{lp-eq19}
\end{equation}
where $\mu_s=0$ for sources at the contact discontinuity. In the argument of the exponential function, the factor 
\begin{equation}
z_D=\sqrt{{4\,D_0\,E^{a-1}}\over {(1-a)\,\alpha}}
\label{lp-eq20}
\end{equation}
is effectively the propagation length within one synchrotron loss time at the contact discontinuity.

\begin{deluxetable}{rrlr}
\tablecolumns{4}
\tablewidth{0pc} 
\tablecaption{Standard parameters used to derive the figures and their alternative values. The numbers 
are not standard in the sense of being typical for young SNR, but rather one specific set that we use 
to demonstrate the typical behavior of the electron distribution and the emission spectra.
$B_{\rm far}$ is the magnetic field strength far from the contact discontinuity. 
The magnetic field at the 
contact discontinuity is a factor $z_0/z_1$ stronger. $z_s$ denotes the location of the electron 
sources, $a$ indicates the energy dependence of the diffusion coefficient, $D\propto
E^a$, and $\eta_{\rm Bohm}$ is the diffusion coefficient at E=1 TeV at the contact discontinuity in 
units of the Bohm diffusion coefficient.}
\tablehead{\multicolumn{3}{c}{standard} &\colhead{alternative}}
\startdata
$z_1$&0.1&pc& \\
$z_0$&10&$z_1$& \\
$E_{\rm max}$&\phm{W}100&TeV\phm{bla}& \\
$B_{\rm far}$&10&$\mu$G& \\
$z_s$&0& &50 $z_1$\\
$a$&0.9& &0.6\\
$\eta_{\rm Bohm}$&2& &20
\enddata
\label{lp-t1}
\end{deluxetable}

\begin{figure}
\plotone{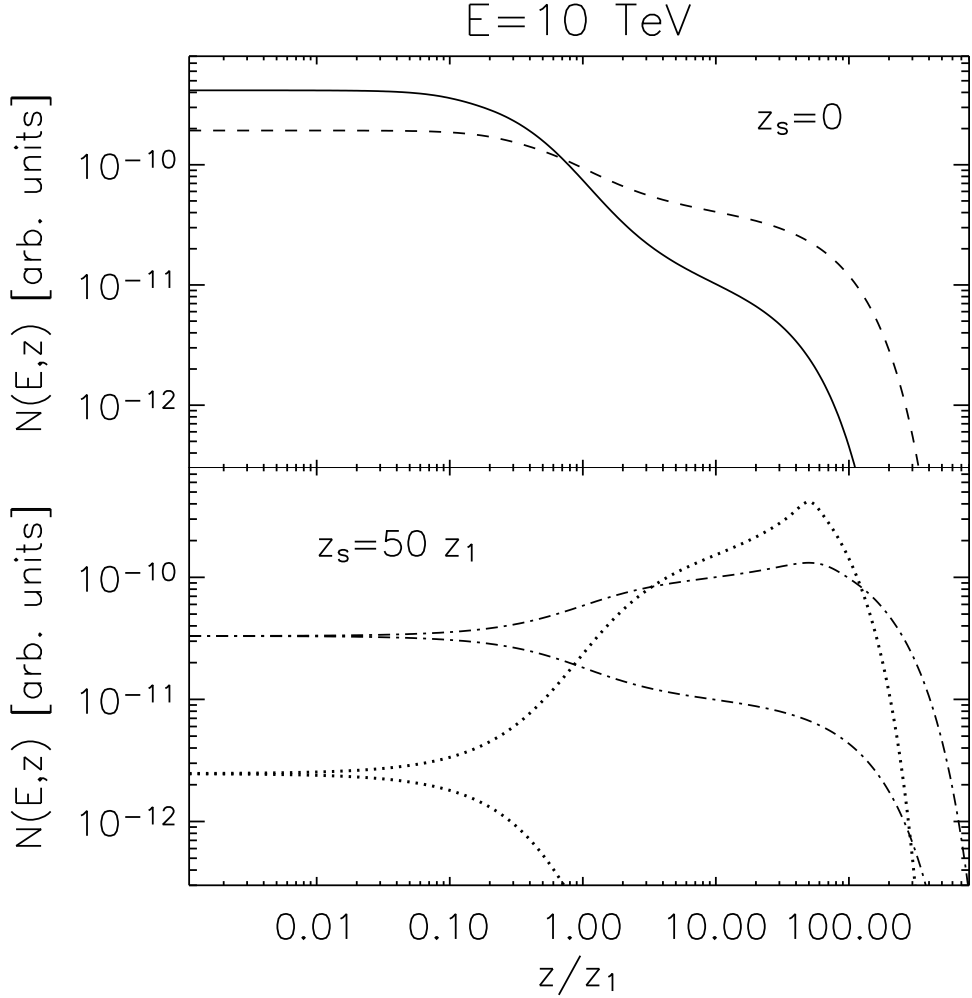}
\caption{ 
The electron flux at $E=10\,$TeV as a function of position for
sources at the contact discontinuity (top panel).
The solid line indicates the electron flux for the standard parameters listed in Table \ref{lp-t1}.
The electron distribution is symmetric around $z=0$ for sources at the contact discontinuity.
For comparison, with the dashed line we also show the electron distribution derived for a 
higher diffusion coefficient ($\eta_{\rm Bohm}=20$) with smaller energy dependence ($a=0.6$). 
The bottom panel shows
the electron flux for sources outside of the contact discontinuity at $z_s=50\,z_1=5$~pc, in 
which case the distribution is not symmetric around $z=0$ and two profiles must be shown, the higher one 
applying to the source side of the contact discontinuity and the lower line to the side far from the 
sources. The dotted lines are for a standard diffusion coefficient and the dash-dotted lines for 
$\eta_{\rm Bohm}=20$, but standard energy dependence $a=0.9$. Obviously the diffusion coefficient must 
be very high to transport a copious supply of electrons to the contact discontinuity, where they can
efficiently radiate.}
\label{lp-f1}
\end{figure}

In Figure \ref{lp-f1} we illustrate the characteristic behavior of the 
differential electron density. For this purpose we use a 
standard set of parameters, shown in Table~\ref{lp-t1}, and compare the spectra thus derived with
those for a non-Bohmian diffusion coefficient and for electron sources far from the contact discontinuity. 
For the standard parameters the propagation length $z_D (1\,{\rm TeV})\simeq 3.4\,z_1$. 
To be noted from the figure is that for sources at the contact discontinuity the width of the spatial 
distribution of the electron flux falls off much closer to the contact discontinuity than indicated 
by the propagation length in one loss time, $z_D$,
on account of the additional factor $1/(1-x^{a-1})$ in the argument of the exponential function
in Eq.\ref{lp-eq19}.
For comparison, as the dashed line we also show the electron distribution derived for a 
higher diffusion coefficient with smaller energy dependence ($a=0.6$), for which the propagation
length is $z_D (1\,{\rm TeV})\simeq 5.4\,z_1$. Though the spatial distribution is clearly wider than for
the standard parameters, it also starts rolling off very close to the contact discontinuity.

If the electron sources are located far from the contact discontinuity, e.g. at the forward shock,
very few electrons will reside at or be transmitted through the contact discontinuity, unless the 
diffusion coefficient is very high. With the dash-dotted line we also show the electron 
distribution derived for a 
higher diffusion coefficient with $\eta_{\rm Bohm}=20$, but standard energy dependence $a=0.9$, 
for which the propagation length is $z_D (1\,{\rm TeV})\simeq 10.7\,z_1$.
Most high-energy electrons would be located near to the source location.

\subsection{The integrated emission spectra}

Equipped with the electron flux as a function of location and energy, we can now proceed to 
calculate the emission properties. We will begin with the volume-integrated spectra at X-ray energies
and in the TeV band. We do not know the electron source strength, $q_0$, and therefore we can not make a 
prediction for the absolute fluxes. However, we can determine the X-ray and $\gamma$-ray fluxes
relative to each other, which should be sufficient for the comparison with the multi-band spectrum of
individual SNRs. In the simple one-zone models, the TeV-to-X-ray flux scaling is a measure of the 
magnetic field strength (Pohl 1996). In our spatially inhomogeneous model the situation is more 
complicated.

In the calculation of the TeV-band $\gamma$-ray emission we will consider only the up-scattering of 
the microwave
background, partly because of the Klein-Nishina cut-off and partly because the ambient
photon densities in the far-infrared will often be too small to significantly influence the results.
In any case, a modeling of the up-scattered infrared emission will require a careful assessment of the
infrared photon fields near and in the SNR in question and thus can be performed only for a specific
object (e.g. see Atoyan et al. 2000). In the isotropic case, 
the differential cross section for the scattering of a
photon with incident energy $\epsilon$ to the energy $E_\gamma$ by an elastic collision with an
electron of energy $E$ is given by (Blumenthal \& Gould 1970)
\begin{equation}
\frac{d\sigma}{d\epsilon} (E_\gamma, \epsilon, E)={{3\,\sigma_T\,m_e^2\,c^4}\over {4\,\epsilon\,E^2}}
\,\left[2q\,\ln q+ (1+2q)(1-q) +{{(1-q)\,(\Gamma_e\,q)^2}\over {2\,(1+\Gamma_e\,q)}}\right]
\label{lp-eq21}
\end{equation}
where
\begin{equation}
q={{E_\gamma}\over {\Gamma_e\,(E-E_\gamma)}}\quad{\rm and}\ \Gamma_e={{4\,\epsilon\,E}\over {m_e^2\,
c^4}}
\end{equation}
In our case $\Gamma_e\approx 1$ and thus the Thomson limit is not valid. Given the differential
electron number density, $N(E)$, the $\gamma$-ray emissivity in the TeV band is calculated as
\begin{equation}
j_\gamma ={{c\,E_\gamma}\over {4\pi}} \,\int_{E_{\rm min}} dE\,\int d\epsilon\ 
n(\epsilon)\,N(E)\,\frac{d\sigma}{dE_\gamma}
\label{lp-eq22}
\end{equation}
where
\begin{equation}
n(\epsilon)= {1\over {\pi^2\,\hbar^3\,c^3}}\,{{\epsilon^2}\over {\exp\left({\epsilon\over {kT}}
\right) -1}}
\label{lp-eq23}
\end{equation}
is the blackbody photon density spectrum of the microwave background (with $T=2.73\,$K).
The total emission spectrum is then calculated as the volume integral of the emissivity, which 
essentially corresponds to using the volume-integrated electron spectrum in Eq.\ref{lp-eq22}.

The corresponding volume integral of the synchrotron emissivity is more complex, for the magnetic
field strength, $B$, varies considerably over the emission region. The volume-integrated X-ray and
TeV-band spectra are shown in Fig.\ref{lp-f2} {both for sources at and far from the contact discontinuity,
using the same parameter combinations as in Fig.\ref{lp-f1}. } 
To be noted from the figure is that by variation of the
propagation length, $z_D$, one can indeed reduce the keV-to-TeV flux ratio, as was suggested by
Allen, Petre \& Gotthelf (2001) and Lazendic et al. (2004). {However, more important
appears the location of the electron sources, for }the smallest flux ratio is produced, if
the sources are far from the contact discontinuity. In that case most of the synchrotron emission
is generated in regions of low magnetic field strength and the X-ray spectrum is 
correspondingly soft.

\begin{figure}
\plotone{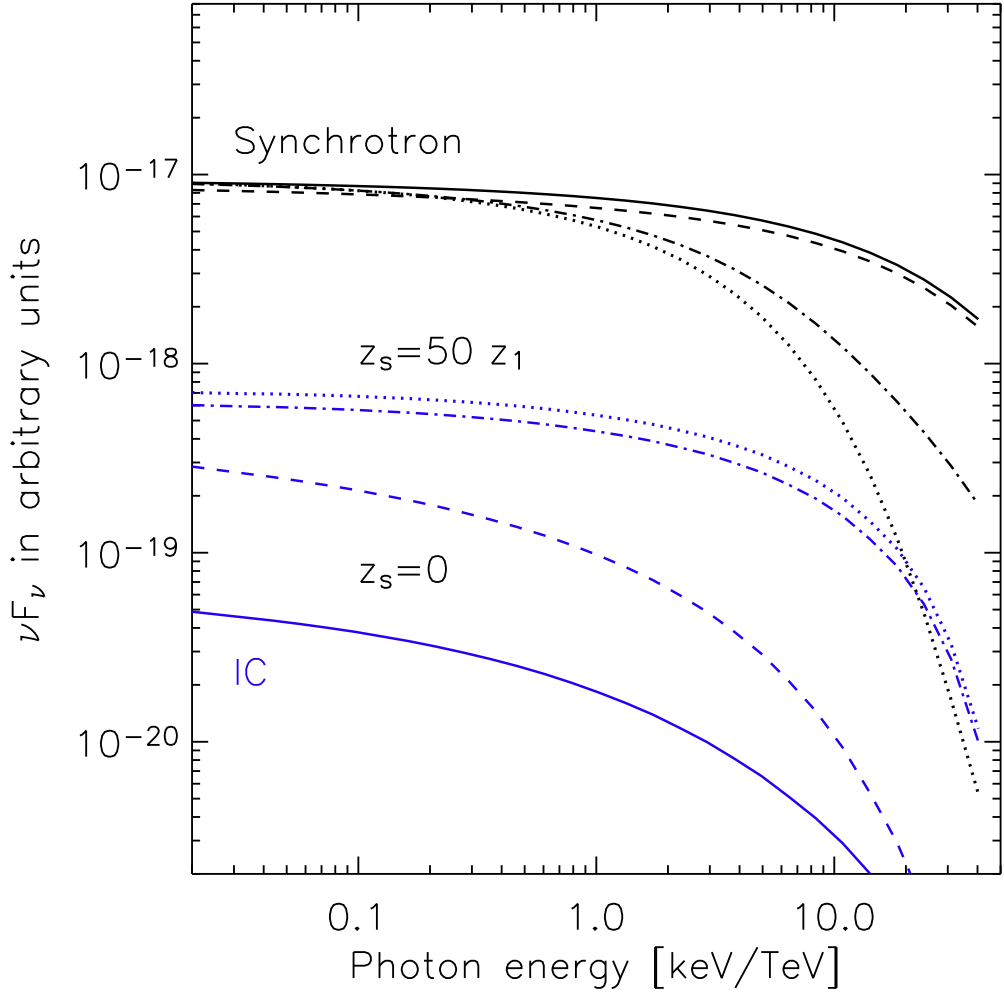}
\caption{The volume-integrated emission spectra at X-ray energies (by synchrotron emission)
and in the TeV band (by inverse Compton scattering). The fluxes are in arbitrary units but obey the
correct relative scaling. The relation between the linestyles and the parameters used are as in 
Fig.\ref{lp-f1}. The abscissa units are keV for the synchrotron emission and TeV for the inverse Compton 
spectra.}
\label{lp-f2}
\end{figure}
Please note that
the spectra shown in Fig.\ref{lp-f2} have been calculated assuming a steady-state situation. While
that appears justified at X-ray energies of a few keV or higher, it may be inappropriate for low-energy
X-rays. Depending on the age and the acceleration history of the supernova remnant in question, there will
be a spectral transition somewhere in the UV or soft X-ray regime from a spectrum at lower energies, that
is essentially unaffected by energy losses and just reflects the source spectrum, 
to a steady-state spectrum at higher energies, that we discuss here. As a consequence, the low-frequency 
synchrotron spectrum will be harder by a spectral index change $\Delta s=0.5$ compared with the
high-frequency spectrum. A corresponding turnover should be observable in the inverse Compton
spectrum at TeV energies. 
This spectral feature arises from the relation between the electron energy loss time and the age
of the system, and it has nothing to do with the usual energy loss limits on the maximum achievable
energy in the electron acceleration process. The typical electron spectrum will have a cut-off and
a spectral break, the former being the consequence of the energy limit of the electron acceleration,
and the latter being caused by the transition from the pure source spectrum at low energies to a 
steady-state situation at higher energies. A radiation modeling of SNRs based on simple 
power-law spectra with cut-off is therefore inappropriate.

\subsection{X-ray filaments: the intensity profile at the contact discontinuity}
While the curvature of the contact discontinuity appeared negligible in the treatment of the electron
propagation, this is clearly not the case when discussing X-ray intensity profiles at the contact discontinuity. The
observable intensity distribution will be different from the spatial profile of
the synchrotron emissivity, $j_X$, as a function of the distance from the SNR center, $r$.
Assuming spherical symmetry of the SNR we can calculate line-of-sight integrals of X-ray intensity
in the energy interval $\left[E_1,E_2\right]$ as
\begin{equation}
I_X (r_0)=\int_{E_1}^{E_2}\,dE_X\ E_X^{-1}\,\int_{-\infty}^\infty dx\ j_X 
\left(E_X, r=\sqrt{r_0^2 +x^2}\right)
\label{lp-eq24}
\end{equation}
where $r_0$ is the distance between the line-of-sight and the center of the remnant.
{The observable intensity distribution (Eq.\ref{lp-eq24}) will depend on
our choice of diffusion coefficient, $\eta_{\rm Bohm}$, source location, $z_s$, and thickness
of the magnetic pile-up region, $z_1$}. To solve this line-of-sight integral
we also need to specify the radius of the contact discontinuity, $r_{\rm CD}$, for in the preceding
calculations we have used a spatial coordinate $z$ with the contact discontinuity as the origin, so that now
$r=r_{\rm CD} +z$. The resulting profiles for the parameters given in Table \ref{lp-t1} and
$r_{\rm CD}=10\,$pc are shown in Fig.\ref{lp-f3} for the typical X-ray energy band 3-10~keV. 
To be noted from the figure is the large 
difference between the intensity scale length upstream and downstream of the contact discontinuity. For the standard 
parameters (the solid line) the downstream scale length is approximately a factor of seven larger
than upstream. The dotted lines refer to electron sources far from the contact discontinuity 
and apparently the magnetic field pile-up does not cause a corresponding enhancement in the X-ray 
intensity. Thus it seems that the high-energy electrons must be produced close to or at the contact discontinuity to
explain the narrow X-ray filaments. 

Also apparent in Fig.\ref{lp-f3} is that extended diffuse emission may be observed far from the 
contact discontinuity, the intensity of which downstream and upstream depends on the source location and the diffusion
coefficient.
 
\begin{figure}
\plotone{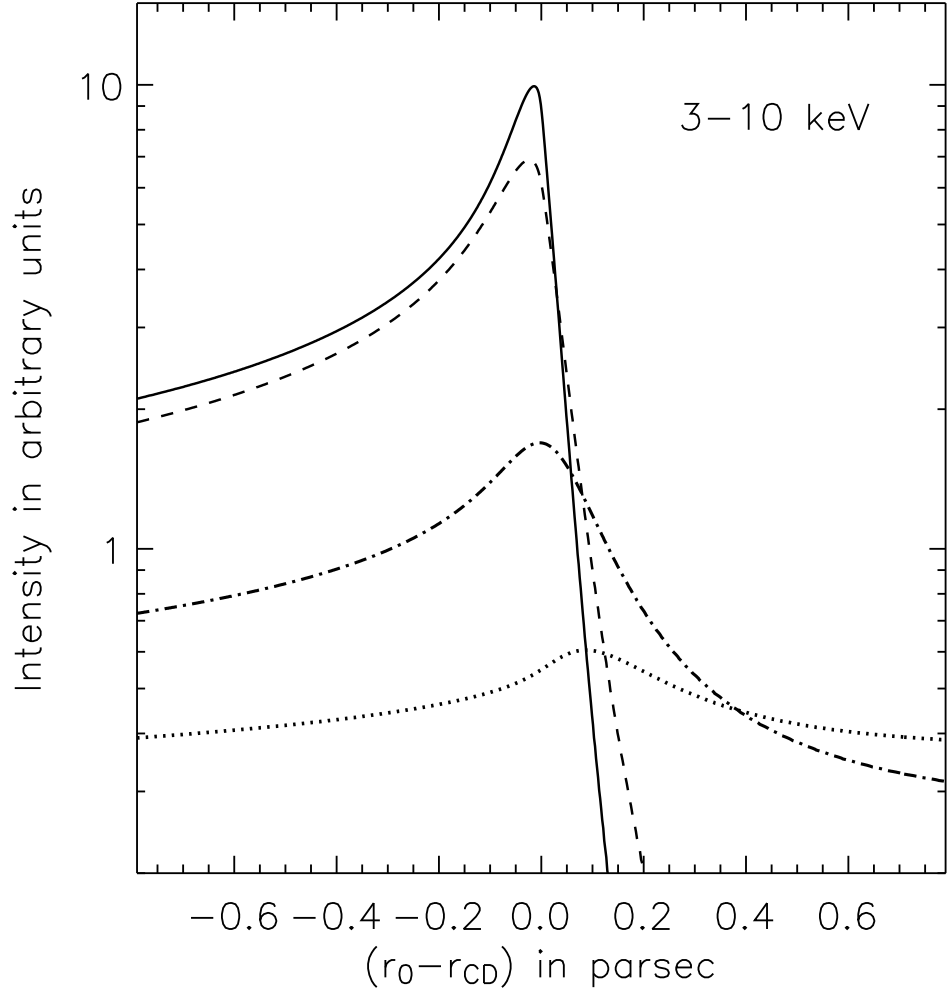}
\caption{The X-ray intensity distribution in the 3-10~keV energy band as seen by an outside observer
assuming spherical symmetry of the SNR (Eq.\ref{lp-eq24}). The different linestyles correspond to 
the same sets of parameters as in Fig.\ref{lp-f1}. Only if the electron sources are close to or at the
contact discontinuity (solid and dashed lines), a narrow enhancement or filament would be observed in the
X-ray intensity profile.}
\label{lp-f3}
\end{figure}

\begin{figure}
\plotone{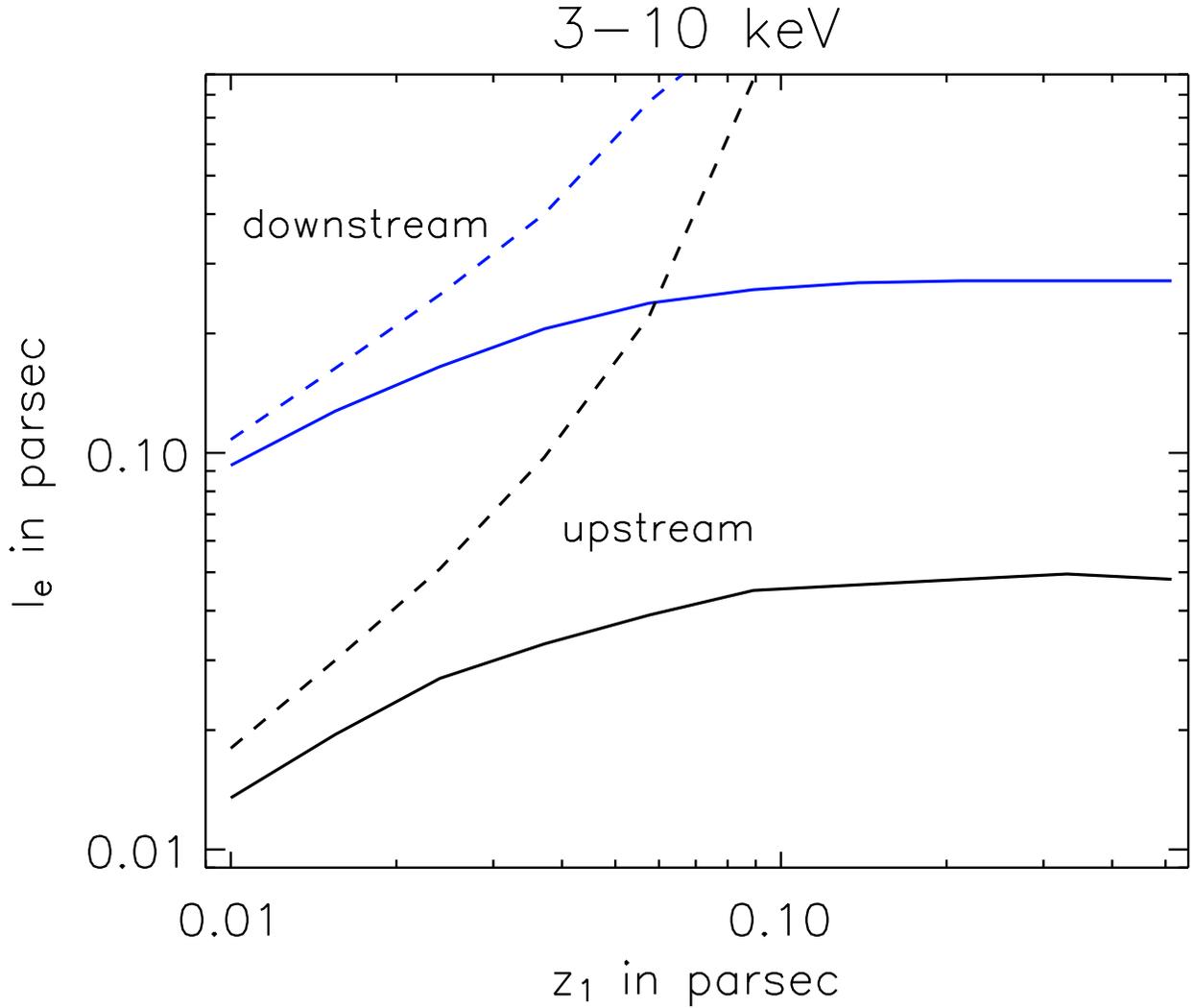}
\caption{The exponential scale length for the 3-10 keV band, $l_e$, upstream and downstream of 
the contact discontinuity, here plotted as a function of the thickness of 
the magnetic field pile-up region, $z_1$. The solid lines refer to the standard values
for all other parameters, whereas the dashed lines are for a source location outside
of the pile-up region at $z_s=20\,z_1$.}
\label{lp-f4}
\end{figure}

\begin{figure}
\plotone{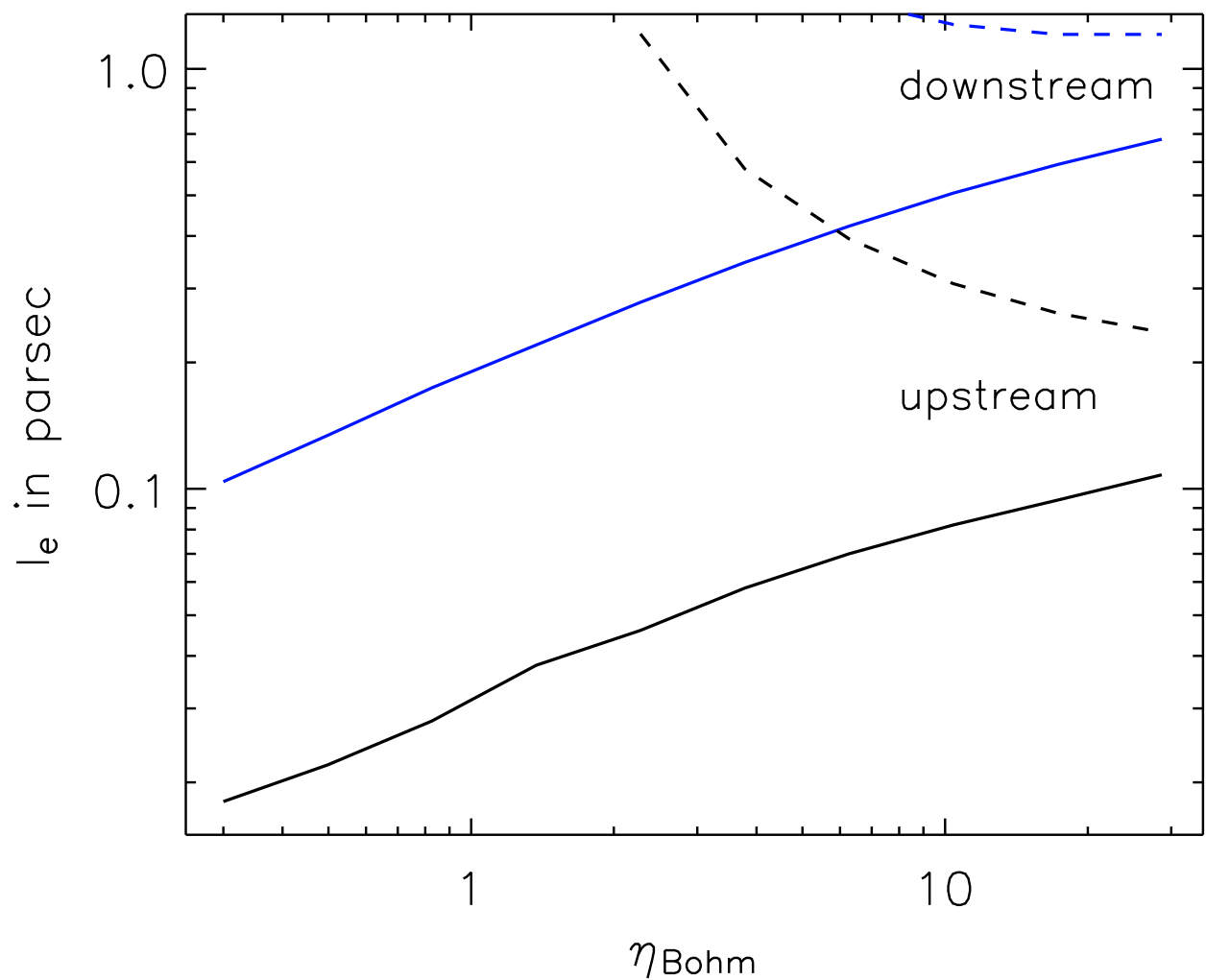}
\caption{The exponential scale length for the 3-10 keV band, $l_e$, displayed as a 
function of the diffusion coefficient
in units of its Bohm limit, $\eta_{\rm Bohm}$. The linestyles and their
relation to the parameters are as in figure \ref{lp-f4}.}
\label{lp-f5}
\end{figure}

\begin{figure}
\plotone{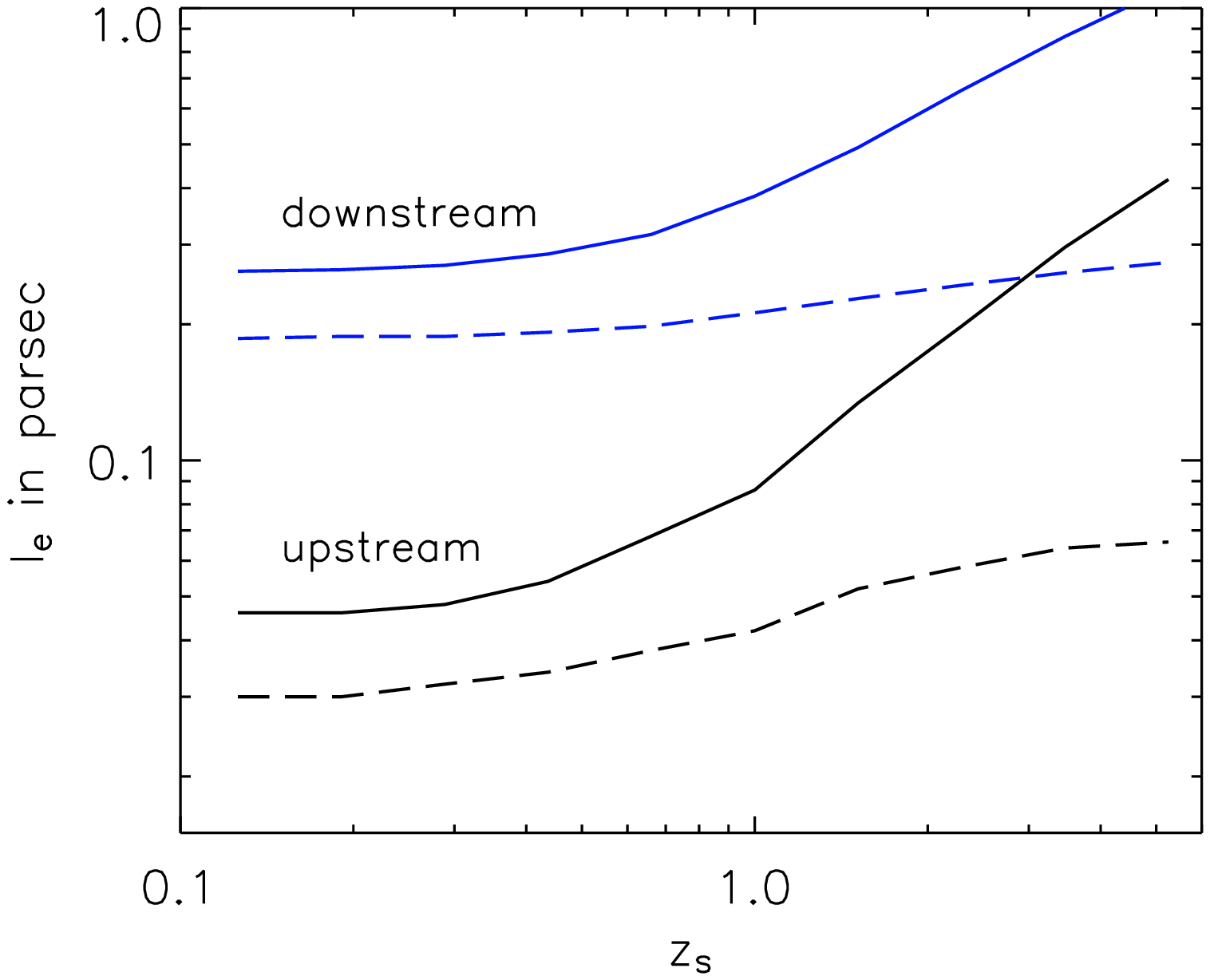}
\caption{The exponential scale length for the 3-10 keV band, $l_e$, shown versus the 
electron source location, $z_s$. Again, the solid lines refer to the standard values
for all other parameters, whereas the dashed lines are for a reduced width of the 
magnetic pile-up region, $z_1=0.03$~pc.}
\label{lp-f6}
\end{figure}

How does the filament width scale with the parameters, such as the width of the pile-up region,
$z_1$, the diffusion coefficient, $\eta_{\rm Bohm}$, or the source location, $z_s$? 
The intensity profile near the contact discontinuity seems to have a variety of shapes, depending on where the electron
sources are located, and thus one measure of filament width doesn't do justice to all situations.
With that limitation in mind, we have calculated the distance from the point, where the intensity
in the 3-10 keV band is 
at its peak value, to the location, where it has fallen to $1/e$ of that, and use the quantity thus
derived as the thickness scale length $l_e$ of the X-ray filament. Obviously $l_e$ will be different 
upstream and downstream of the contact discontinuity on account of its spherical geometry and possibly the electron
source location. 

In the figures
\ref{lp-f4}, \ref{lp-f5}, and \ref{lp-f6} we show the thickness scale length, $l_e$, as a function of
the thickness of the magnetic field pile-up region, $z_1$, in dependence of the diffusion coefficient, 
$\eta_{\rm Bohm}$, and versus the electron source location, $z_s$, respectively, while in each 
case all the other parameters have been kept at their standard values as listed in table \ref{lp-t1}.

To be noted from Fig.\ref{lp-f4} is that very thin X-ray filaments can be produced
{whatever the source location}, if the thickness
of the magnetic field pile-up region, $z_1$, is very small, though much less magnetic flux is
compressed at the contact discontinuity in that case. However, {only for electron sources at the contact discontinuity
the thickness of the X-ray filaments remains small even for
high values of $z_1$. The scale length, $l_e$, turns constant for large $z_1$, because the 
energy losses do no longer permit the 
high-energy electrons to fully occupy the region of enhanced magnetic field. For $z_1\gtrsim
0.03$~pc narrow filaments will not be observed, if the electron sources are far from the contact discontinuity.}

If one increases the diffusion coefficient from values below the Bohm limit, the filament thickness 
will also increase. However, Fig.\ref{lp-f5} indicates that the increment in the scale length
$l_e$ is less than that in the propagation length scale $z_D$($\propto \sqrt{\eta_{\rm Bohm}}$).
Thus, both the thickness of the magnetic pile-up region and the electron diffusion coefficient 
must be sufficiently small to produce narrow X-ray filaments, if the radiating electrons 
are produced at the contact discontinuity. {For electron sources far from the contact discontinuity (indicated by the dashed line),
the diffusion coefficient must be large enough to allow a significant fraction of the electrons
to propagate to the high magnetic field region around the contact discontinuity, otherwise narrow filaments
will not be observed.}

Fig.\ref{lp-f6} shows that if the electron sources are not located in the magnetic pile-up region
around the contact discontinuity anymore ($z_s > 1$), two things will happen {unless the thickness
of the pile-up region is very small (this case is indicated by the dashed lines)}: 
firstly, the width of the X-ray filaments will substantially increase, in fact approximately 
linearly with $z_s$ in case of the upstream length scale. Secondly, whereas for $z_s\approx 0$ 
the intensity length scale $l_e$ downstream is roughly a factor of 5 larger than that upstream 
of the contact discontinuity, the two length scales tend to get closer to each other with
increasing $z_s$. 

{Summarizing our results on the X-ray scale length we note that if the electron sources are
located far from the contact discontinuity, only for a sufficiently thin magnetic pile-up region with 
$z_1\lesssim 0.03$~pc can narrow X-ray filaments with a downstream exponential length scale 
$l_e\lesssim 0.3$~pc be observed. On the other hand, if the electrons are accelerated at the
contact discontinuity, narrow X-ray filaments would be produced for a broad variety of diffusion coefficients
and widths of the magnetic field sheath. We therefore conclude that if the X-ray filaments 
are caused by magnetic field pile-up at the contact discontinuity, it is much more likely that the relativistic
electrons are locally accelerated than far from the contact discontinuity.}

We have also searched for spectral differences between the X-ray intensity of the filaments and that
of the extended downstream regions. For that purpose we have defined the filament region as the area, 
in which the 3-10 keV band integrated intensity is higher than half of its maximum. The downstream 
region then extends from the filament to 1 pc away from the contact discontinuity. The filament intensity and plateau
intensity are just averages over the areas in question,
\begin{eqnarray}
I_{\rm fil}&=&{1\over {l_{1/2}^+ + l_{1/2}^-}}\,\int_{r_{CD}-l_{1/2}^-}^{r_{CD}+l_{1/2}^+} 
dr_0\ I_X (r_0) \\
I_{\rm pla}&=&{1\over {[1\,{\rm pc}] - l_{1/2}^-}}\,\int_{r_{CD} - [1\,{\rm pc}]}^{r_{CD}-l_{1/2}^-} 
dr_0\ I_X (r_0)\label{lp-eq25}
\end{eqnarray}
electron \begin{figure}
\plotone{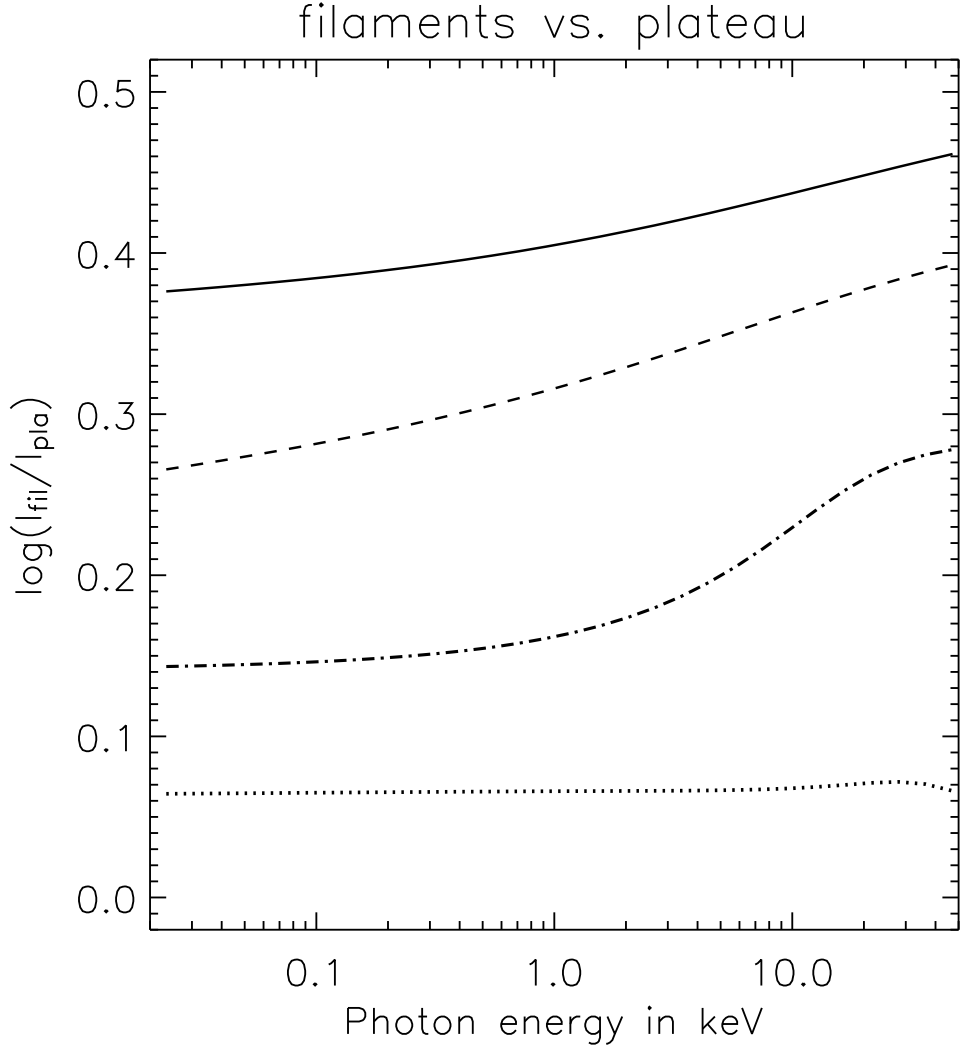}
\caption{The ratio of the average X-ray intensities of the filaments and the extended downstream regions
according to Eq.\ref{lp-eq25}, shown as a function of the photon energy.
Here 'filament' is defined by the X-ray intensity in the 3-10 keV band being at least 50\% 
of the peak value. The downstream spectrum is derived in the region between the filament and 
1 pc away from the contact discontinuity, i.e. a bit further than what is shown in Fig.\ref{lp-f3}.
The filament spectra seem to be slightly harder than those of the downstream region, if with a 
spectral index difference of $\Delta \alpha \lesssim 0.05$ probably too little to be detectable.
The relation between the linestyles and the parameters used are as in Fig.\ref{lp-f1}.}
\label{lp-f7}
\end{figure}
In Fig.\ref{lp-f7} we show the ratio of these intensity spectra for the same four sets of parameters
that we have used in the Figures \ref{lp-f1}-\ref{lp-f3}. If the spectral form of the emission from the
filaments and the plateau regions was the same, the lines would be flat at a value defined by the jump
in the mean intensity. Excluding the case of sources far away and a small diffusion coefficient, in which
filaments are only marginally existent, all curves have a positive slope, thus indicating that the 
filament spectra seem to be slightly harder than those of the downstream region. However, the difference
in spectral index is with $\Delta \alpha \lesssim 0.05$ probably too small to be observable.{
In the case of RX~J1713.7-3946 claims of spectral constancy (Uchiyama \etal 2003) have been published.
Care must be exercized, though, for there is a clear ambiguity in spectral fits related to the 
absorbing column density and the 
emission spectral index. The spectral constancy reported by Uchiyama \etal (2003) requires that the absorbing
column is highest toward the regions of highest intensity, which seems unlikely. Keeping the absorbing column fixed
in the spectral analysis, Lazendic \etal (2004) find the X-ray spectra of the filaments harder than those of
the plateau emission.  
}
\section{Magnetic dissipation at the contact discontinuity?}

Can dissipation of the magnetic field power the non-thermal X-ray emission? 
If we estimate the  total X-ray luminosity of a typical young SNR to be
$10 ^{35}-10^{36}$~erg/sec and a typical age $10^3-10^4$~yrs, then the total emitted
energy over the life time  is $\lesssim 10^{47}$~ergs, or about $0.01\%$ of the total mechanical
energy of a SNR, $E_{\rm tot}$. The energy in the non-thermal component should be smaller 
than the total free energy available for particle acceleration.

The total energy associated with the magnetic field in the sheath can be estimated 
as 
\be
E_{sheath} \sim {\Delta R _{sheath} \over R} E_{tot} \sim {  E_{tot}  \over M_A^{9/4}}
 \sim \left(10^{-2} -  10^{-3}\right) \, E_{tot} \sim 10^{48} {\rm ergs},
\label{Es}
\ee
This estimate (\ref{Es}) implies  that the total X-ray energetics 
of SNRs is only slightly less than the total magnetic energy of the swept-up material.
In addition, the magnetic field of the ejecta will contribute to the 
nonthermal emission as well, but its contribution cannot be estimated due to the unknown
values of the magnetic field in the ejecta. \footnote{There is  a 
renewed interest in magnetically-driven SN explosion, stimulated by 
the association of some SN with GRBs.}

One possible mechanism {for the energy transfer from a magnetic field to relativistic particles}
is magnetic reconnection in the compression region at the contact discontinuity. 
Note, that  the contact discontinuity is, in fact, a
 rotational discontinuity of magnetic field, so
that there are current flowing along it. Dissipation of those currents may, in principle, 
power the  non-thermal emission.
Magnetic reconnection leads to a large-scale DC electric field
which efficiently  accelerate
particles either directly, or through generation of MHD turbulence
(\eg Priest \& Forbes 2000).  
The accelerating electric field  can be  parameterized as
 $E = \xi B$, $\xi \leq 1$ (this is 
equivalent to acceleration on a time scale $t_{acc} \sim \xi \gamma /\om_B$; in a framework
of reconnection models (\eg, Priest \& Forbes 2000)
 $ \xi = v_{in}/c$ where $v_{in}$ is the inflow velocity). 
Balancing the acceleration rate with the
synchrotron energy loss rate, {we find for the maximum Lorentz factor, $\gamma_{\rm max}$,
and the energy of the corresponding synchrotron photons, $E_{\rm max}$,}
\ba &&
E_{X,\,\rm max} = \xi \, \frac{27}{8}\,{ {m\,c^2} \over \alpha}
 \nn &&
\gamma_{\rm max} = \frac{3}{2}\,\sqrt{ \xi\, { {m\,c^3} \over {e^2\, \om_B}}}
\ea
where $\alpha$ is the fine structure constant.
Then, in order to have the observed X-ray synchrotron emission in the
$\sim 5 $~keV range one requires
\be
 \xi >  \alpha \, \frac{8}{27}\,{ E_{X} \over {m\,c^2}} \simeq  2\cdot 10^{-5} \,
\left({{E_{X}}\over {5\ {\rm keV}}}\right) 
\ee
{There will be other loss processes, e.g. associated with scattering on magnetic turbulence,
that will limit the efficacy of acceleration and hence may require a value of $\xi$ 
larger than calculated here. Nevertheless, values of $\xi \ll 1$ may be sufficient to accelerate
electrons such that they can emit X-ray synchrotron radiation.}

The treatment of magnetic dissipation is notoriously complicated and has little predictive 
power {(see, e.g.,
Lesch \& Pohl 1992). For example on cannot calculate
$\xi$ from first principles. In this section our intention was to point out the
consistency of basic back-of-the-envelope estimates for the total energetics
and the acceleration rates with the notion that 
non-thermal X-ray emitting  electrons may be accelerated at the contact discontinuity.}

\section{Discussion}

We have investigated the possibility that the non-thermal X-ray filaments observed in pre-Sedov
SNRs are related to the dynamical compression
of magnetic field at the contact discontinuity that separates the circumstellar medium and the ejecta.
{The pile-up of magnetic field at the contact discontinuity has been extensively studied in the literature 
(Kulsrud \etal 1965; Lee \& Chen 1965; Rosenau 1975; Rosenau \& Frankenthal 1976).
We have combined the magnetohydrodynamical structure of magnetized outflows as described in these
publications with a one-dimensional diffusion-convection transport equation 
to describe the propagation of non-thermal electrons near the contact discontinuity of a young
SNR and to calculate spatially resolved emission spectra in the X-ray and TeV bands. 
We have considered two possible acceleration regions: one in the low field region far from the contact discontinuity, 
and the other at the contact discontinuity.

As a result we find that the X-ray intensity distribution will depend strongly on the electron
diffusion coefficient and on the thickness of the magnetized sheath at the contact discontinuity.
If the electron sources are
located far from the contact discontinuity, only for a sufficiently thin magnetic pile-up region with thickness
$z_1\lesssim 0.03$~pc can narrow X-ray filaments with a downstream exponential length scale 
$l_e\lesssim 0.3$~pc be observed. On the other hand, if the electrons are accelerated at the
contact discontinuity, narrow X-ray filaments would be produced for a broad variety of diffusion coefficients
and widths of the magnetic field sheath. We therefore conclude that if the X-ray filaments 
are caused by magnetic field pile-up at the contact discontinuity, it is much more likely that the relativistic
electrons are locally accelerated than far from the contact discontinuity.

We have also calculated the volume-integrated X-ray synchrotron and TeV-band inverse Compton spectra
with a view to thoroughly investigate the expected $gamma$-ray flux. In the case
of RX~J1713.7-3946, measurements taken with the
CANGAROO experiment suggest a TeV-band spectrum (Enomoto \etal 2002), that cannot be readily understood
as inverse Compton emission, but also not as hadronic in origin (Reimer \& Pohl 2002). If the
$\gamma$-ray emission volume was very much larger than the X-ray emission volume, the
inverse-Compton interpretation of the TeV-band spectrum would again be tenable (\eg Pannuti \etal 2003).

Depending on the location of the electron acceleration site, as well as on the diffusion coefficient
and the thickness of the magnetized sheath at the contact discontinuity, the TeV-to-keV flux ratio can vary by more than
an order of magnitude. A relatively high TeV flux is observed either if the diffusion coefficient is
very large or if the radiating electrons are accelerated far from the magnetic pile-up region, \eg at
the forward shock. These situations are exactly those, for which narrow X-ray filaments can be
observed only if the magnetized sheath at the contact discontinuity is very thin with $z_1\lesssim 0.03$~pc.
Thus, the conditions favorable for the occurrence of thin X-ray filaments have little overlap with
those causing a relatively high TeV flux. We have assumed a magnetic field strength 
$B=10\ {\rm \mu G}$ far from the contact discontinuity, and find that even under favorable circumstances the TeV energy flux
is still an order of magnitude lower than the keV energy flux, which is insufficient to explain
the TeV-band spectrum of RX~J1713.7-3946 observed with CANGAROO. The magnetic field is already compressed
by the forward shock, and a lower field strength, that in principle may reduce the discrepancy, seems
therefore unlikely. Furthermore, in cases of high TeV flux the TeV-band spectrum is typically hard in 
comparison with the X-ray spectrum, also in contrast to the measurements. We therefore feel,
that our model does not offer a comprehensive explanation for the CANGAROO measurements of
RX~J1713.7-3946.}

Dynamical compression of magnetic field at the contact discontinuity generically happens 
in the  pre-Sedov stage, which may explain why non-thermal filaments appear
both in Type II and in Type Ia supernova. The particular appearance of filaments and their
relative importance in producing X-rays may depend on the 
progenitor. Core  collapse SN are preceded by several
stages of wind-related mass loss, during which a considerable amount of magnetic flux is expelled.
After the SN explosion the wind's magnetic field is then compressed at the contact discontinuity.
This is applicable to Cas~A, and RX~J1713.7-3946.
On the other hand, SN~1006 was probably Type Ia event (Winkler \etal 2003) and thus is not expected 
to have a considerable wind-blown nebular around it.
Explosions into the interstellar medium are expected to produce very thin 
magnetized sheaths, with a thickness
$\Delta R/R \sim M_A^{-3}$ (Kulsrud \etal 1965),
so that only a fraction $M_A^{-2}$ of the swept-up
flux is concentrated at the contact discontinuity.
In this case, the magnetic flux from the SN progenitor itself,
dynamically compressed at the {\it inner} side  of the contact discontinuity, {may be more important than 
the field of the circumstellar material (\eg Lou 1994).}

One of the arguments in favor of a forward-shock interpretation
of the filaments in SN~1006 is that their position coincides with the $H_\alpha$ rim.
$H_\alpha$ emission often  occurs at shock fronts due to both charge exchange reactions and
collisional heating. One may expect that, 
if particles are accelerated at the contact discontinuity, this
should be accompanied by plasma heating, similar to Solar flares. 
This, in principle, can produce $H_\alpha$ emission as well if there is 
atomic hydrogen present near the contact discontinuity.

As the SNR becomes older, it evolves from the ejecta-dominated to the Sedov-dominated
phase. During this evolution the contact discontinuity slows down its motion, receding further
from the forward shock, and is finally pushed back to smaller radii (\eg Truelove \& McKee 1999).
As the contact discontinuity recedes from the shock, smaller 
and smaller amounts of the magnetic flux
are concentrated on it.
In addition, if the forward shock becomes radiative, the amount of the magnetic flux
piled on the contact discontinuity decreases even further, 
as radiative shock have larger compression ratios,
so that less flux is advected downstream.
Thus, the non-thermal emission will subside after the SNRs either enter the Sedov phase
or the forward shock becomes radiative.

The location of the contact discontinuity should vary as the expanding SNR collides with inhomogeneities
{caused by the progenitor's wind or structure in the interstellar medium}, 
thus producing a complicated structure. In addition,
the contact discontinuity may be unstable with respect to Rayleigh-Taylor and
Richtmyer-Meshkov processes (Kane \etal 1999). A stretching
of the toroidal magnetic field by density clumps and by instabilities
may explain why radio polarization studies indicate that projected large-scale magnetic
fields tend to be oriented in the radial direction (\eg Reynolds \& Gilmore 1993).

\section{Acknowledgements}
We would like to thank  Felix Aharonyan, Roger Blandford, Roger Chevalier,  Steve Reynolds
and Chris Thompson.
ML acknowledges support by  NSERC grant  RGPIN 238487-01. 
MP acknowledges support by NASA under award No. NAG5-13559.


\begin{thebibliography}{}

\bibitem{2001} Aharonian, F. A. \etal. 2001a, A\&A, 370, 112

\bibitem{2001b} Aharonian, F. A. \etal. 2001b, A\&A, 373, 292

\bibitem{2002} Aharonian, F. A. \etal. 2002, A\&A, 395, 803

\bibitem{all01} Allen, G.E., Petre, R., and Gotthelf, E.V., 2001, \apj, 558, 739

\bibitem{Allen97} {Allen}, G.~E. \etal, 1997,
{\apjl}, 487, 97

\bibitem{at2000} Atoyan, A.M., Aharonian, F.A., Tuffs, R.J., and V\"olk, H.J., 2000, \aap, 355, 211

\bibitem{Bam00} {{Bamba}, A., {Koyama}, K., and {Tomida}, H.},  2000, {\pasj}, 52, 1157

\bibitem{Bam03}
{{Bamba}, A., {Yamazaki}, R., {Ueno}, M., {Koyama}, K.},
2003, {\apj}, 589, 827

\bibitem{bl01} Bell, A.R. \& Lucek, S.G., 2001, \mnras, 321, 433

\bibitem{Berezhko02} {{Berezhko}, E.G, {Ksenofontov}, L.T., and {V{\" o}lk}, H.J.
        }, 2002, \aap, 395, 943

\bibitem{Berezhko03} {{Berezhko}, E.G., {P{\" u}hlhofer}, G., and {V{\" o}lk}, H.J.
        },     2003,  {\aap}, 400, 971

\bibitem{Berezhko03b} {{Berezhko}, E.G., {Ksenofontov}, L.T., and {V{\" o}lk}, H.J.
        }, 2003, \aap, 412, L11

\bibitem{bg70} Blumenthal, G.R. \& Gould, R.J., 1970, Rev. Mod. Phys., 42-2, 237

\bibitem{Borkowski01} {{Borkowski}, K.J., {Rho}, J., {Reynolds}, S.P., and {Dyer}, K.K.
        }, 2001, {\apj},  550, 334

\bibitem{bu98} Buckley, J.H., et al., 1998, \aap, 329, 639

\bibitem{cas01} {{Cassinelli}, J.P.},
2001, in {\it Observable Effects of B Fields on the Winds and Envelopes of Hot Stars},
ASP Conf. Ser. 248, Mathys, G.,   Solanki,  S. K., Wickramasinghe, D. T., Eds,
651

\bibitem{Chevalier94} {{Chevalier}, R.~A. \& {Luo}, D.}, 1994, 421, 225

\bibitem{Chevalier03} {{Chevalier}, R.~A. \& {Oishi}, J.}, 2003, \apj, 593, L23

\bibitem{EmmeringChevalier87} {{Emmering}, R.~T. \& {Chevalier}, R.~A.}, 1987, {\apj},
321, 334 

\bibitem{Enomoto02}   Enomoto, R. \etal, 2002, Nature, 416, 823

\bibitem{Got01} {{Gotthelf}, E.V., {Koralesky}, B., {Rudnick}, L.,
        {Jones}, T.W., {Hwang}, U., and {Petre}, R.},
2001, {\apjl}, 552, 39

\bibitem{Hamilton84} {{Hamilton}, A.J.S. \& {Sarazin}, C.L.}, 1984, {\apj}, 281, 682

\bibitem{kane99} {{Kane}, J., {Drake}, R.P., and {Remington}, B.A.},
1999, {\apj}, 511,
335

\bibitem{Koyama95}  {{Koyama}, K., {Petre}, R., {Gotthelf}, E.V., {Hwang}, U.,
        {Matsuura}, M., {Ozaki}, M., and {Holt}, S.S.}, 1995, {\nat}, 378, 255

\bibitem{Koyama97} {Koyama}, K. \etal, 1997, {\pasj}, 49, 7

\bibitem{Kulsrud}
Kulsrud, R.M., Bernstein, I.B., Krusdal, M., Fanucci, J. and Ness, N., 1965,
\apj, 142, 491

\bibitem[Lazendic \etal(2003)]{la03} Lazendic, J.S., Slane, P.O., Gaensler, B.M., et al., 2004,
\apj, in press, astro-ph/0310696

\bibitem{LeeChen}
Lee, T.S. \& Chen, T., 1965, Planet. Space Sci, 16, 1483

\bibitem{ls81} Lerche, I., Schlickeiser, R., 1981, \apjs, 47, 33

\bibitem{ls82} Lerche, I., Schlickeiser, R., 1982, \aap, 107, 148

\bibitem{lp92} Lesch, H. \& Pohl, M., 1992, \aap, 254, 29

\bibitem{lou94} {{Lou}, Y.},
1994, {\apjl}, 428,
21

\bibitem{lb00} Lucek, S.G. \& Bell, A.R., 2000, \mnras, 314, 65

\bibitem{lyut02} Lyutikov, M., 2002,  Physics of Fluids,  14, 963

\bibitem{murai00} Muraishi, H., Tanimori, T., Yanagita, S., \etal, 2000, \aap, 354, L57

\bibitem[Owens \& Jokipii(1977)]{oj77} Owens, A.J. \& Jokipii, J.R., 1977, \apj, 215, 685

\bibitem{Pan03}
{{Pannuti}, T.G., {Allen}, G.E., {Houck}, J.C., and {Sturner}, S.J.
        }, 2003, \apj, 593, 377

\bibitem{po96} Pohl, M., 1996, \aap, 307, L57

\bibitem{po90} Pohl, M. \& Schlickeiser, R., 1990, \aap, 234, 147

\bibitem{pf00}
{{Priest}, E. \& {Forbes}, T.}, 2000, {\it Magnetic reconnection : MHD theory and applications}, 
Cambridge University Press

\bibitem{rey93} Reynolds, S.P. \& Gilmore, D.M., 1993, \aj, 106, 272

\bibitem{rho02} {{Rho}, J., {Dyer}, K.K., {Borkowski}, K.J., and {Reynolds}, S.P.
        }, 2002, \apj, 581, 1116

\bibitem{rosenau75} Rosenau, P., 1975, Ph.D. Thesis, Tel Aviv University, Ramat Aviv, Israel

\bibitem{Rosenau76}
Rosenau, P. \& Frankenthal, S., 1976, Phys. Fluids, 19, 1889

\bibitem[Skilling(1975a)]{sk75a} Skilling, J., 1975, \mnras, 172, 557
\bibitem[Skilling(1975b)]{sk75b} Skilling, J., 1975, \mnras, 173, 245
\bibitem[Skilling(1975c)]{sk75c} Skilling, J., 1975, \mnras, 173, 255

\bibitem{Slane01} {{Slane}, P., {Hughes}, J.P., {Edgar}, R.J., {Plucinsky}, P.P.,
        {Miyata}, E., {Tsunemi}, H., and {Aschenbach}, B.}, 2001,  {\apj}, 548, 814

\bibitem{slane99} Slane, P., et al., 1999, \apj, 525, 357

\bibitem{Tanimori98} Tanimori, T., et al., 1998,
{\apjl}, 497, 25

\bibitem{Truelove99} {{Truelove}, J.K. \& {McKee}, C.F.}, 1999, {\apjs}, 120, 299

\bibitem{uchi03} Uchiyama, Y., Aharonian, F.A., Takahashi, T., 2003, \aap, 400, 567

\bibitem{Vink03} {{Vink}, J.}, 2004, Adv. Sp. Res., in press, astro-ph/0304176

\bibitem{vink03}{{Vink}, J. \& {Laming}, J.M.}, 2003,
{\apj}, 584, 758

\bibitem[Wentzel(1974)]{we74} Wentzel, D.G., 1974, \araa, 12, 71

\bibitem{wink03} Winkler, P.F., Gupta, G., and Long, K.S., 2003, \apj, 585,324

\end{thebibliography}
\end{document}